# Chaotifying Continuous-Time Nonlinear Autonomous Systems

Simin Yu[1]   and   Guanrong Chen[2]

[1]College of Automation, Guangdong University of Technology, Guangzhou 510006, China

[2] Department of Electronic Engineering, City University of Hong Kong, Hong Kong SAR, China

**Abstract.** Based on the principle of chaotification for continuous-time autonomous systems, which relies on two basic properties of chaos, i.e., globally bounded with necessary positive-zero-negative Lyapunov exponents, this paper derives a feasible and unified chaotification method of designing a general chaotic continuous-time autonomous nonlinear system. For a system consisting of a linear and a nonlinear subsystem, chaotification is achieved using separation of state variables, which decomposes the system into two open-loop subsystems interacting through mutual feedback resulting in an overall closed-loop nonlinear feedback system. Under the condition that the nonlinear feedback control output is uniformly bounded where the nonlinear function is of bounded-input/bounded-output, it is proved that the resulting system is chaotic in the sense of being globally bounded with a required placement of Lyapunov exponents. Several numerical examples are given to verify the effectiveness of the theoretical design. Since linear systems are special cases of nonlinear systems, the new method is also applicable to linear systems in general.

## 1. Introduction

Chaotification, or anti-control of chaos, refers to the task of generating chaos from an originally non-chaotic system by using a simple control input. For continuous-time dynamical systems, several successful techniques have been developed for the task, such as time-delay feedback, impulsive control and topological conjugate mapping [1-11], most of which use a trial-and-error approach to achieve the intended chaotification. In other words, there is no universal and effective framework available in the literature today, except using parameter tuning, numerical simulation and Lyapunov exponent calculation [12-14].

In general, chaotification of continuous-time autonomous dynamical systems starts from one of the two typical settings:

(i)   a linear system $\dot{x} = Ax$ which, when equipped with a controller, results in

$$\dot{x} = A(x) + BG(\sigma x, \varepsilon);$$

(ii) a nonlinear system $\dot{x} = F(x)$, which similarly gives $\dot{x} = F(x) + BG(\sigma x, \varepsilon)$.

Here, $x \in R^n$ is the state vector, $A$ is the system matrix, $F(x)$ is a nonlinear vector-valued function, $G(\sigma x, \varepsilon)$ is a simple nonlinear feedback controller, $B$ is a control matrix, $\sigma$ is a control gain matrix, and $\varepsilon$ is an upper bound for the controller. A standard problem is, for a given $A$ or $F(x)$, design $G(\sigma x, \varepsilon)$, $B$, $\sigma$ and $\varepsilon$, such that the controlled system becomes chaotic.

For linear systems, i.e., case (i) above, the given uncontrolled system needs to satisfy only two simple requirements: the origin of the system is asymptotically stable and the output of the nonlinear controller is uniformly bounded. Under these two conditions, a simple nonlinear state-feedback controller can be designed, such that the controlled system is globally bounded with pre-assigned positive-zero-negative Lyapunov exponents. For nonlinear systems, i.e., case (ii) above, the design is somewhat more difficult and involved. The main challenge lies in the fact that, for nonlinear systems, even if the eiganvalues of their Jacobians at equilibria are all located on the left-half complex plane, a uniformly bounded nonlinear controller may not be able to ensure the controlled system to be globally bounded. To overcome this difficulty, for a system consisting of a linear and a nonlinear subsystem, this paper develops a general chaotification method using separation of state variables to decompose the system into two open-loop subsystems which are interacted via mutual feedback resulting in an overall closed-loop nonlinear feedback system. Under the condition that the nonlinear feedback control output is uniformly bounded where the nonlinear function is of bounded-input/bounded-output, it is proved that the resulting system is chaotic in the sense of being globally bounded with a required placement of Lyapunov exponents.

The rest of the paper is organized as follows. Section 2 introduces the problem description. Section 3 provides the chaotification method and anti-controller design principles. Section 4 demonstrates several representative examples. Finally, Section 5 summarizes the investigation.

## 2. Problem Description

Consider an $n$-dimensional continuous-time autonomous system

$$\dot{x} = F(x) \tag{1}$$

where $x = [x_1, x_2, \cdots, x_n]^T$ is the state vector and $F(x)$ is a nonlinear vector-valued system function of the form

$$F(x) = \begin{pmatrix} F_1(x_1,...,x_n) \\ F_2(x_1,...,x_n) \\ \vdots \\ F_n(x_1,...,x_n) \end{pmatrix} \tag{2}$$

In modern control theory, a basic technical problem is, assuming that a concerned equilibrium of the uncontrolled system (1)-(2) is unstable, to design a nonlinear feedback controller for the system such that the equilibrium becomes asymptotically stable.

On the contrary, a basic problem of chaotification (or anti-control) theory is: Assuming that all the equilibria of the uncontrolled system (1)-(2) are asymptotically stable, design a nonlinear feedback controller $G(\sigma x, \varepsilon)$, as simple as possible, such that the controlled system

$$\dot{x} = F(x) + BG(\sigma x, \varepsilon) \tag{3}$$

becomes chaotic, where $B$ is a control matrix to be designed:

$$B = \begin{pmatrix} b_{11} & b_{12} & \cdots & b_{1n} \\ b_{21} & b_{22} & \cdots & b_{2n} \\ \vdots & \vdots & \ddots & \vdots \\ b_{n1} & b_{n2} & \cdots & b_{nn} \end{pmatrix} \tag{4}$$

and the nonlinear feedback controller

$$G(\sigma x, \varepsilon) = \begin{pmatrix} g_1(\sigma_1 x_1, \varepsilon_1) \\ g_2(\sigma_2 x_2, \varepsilon_2) \\ \vdots \\ g_n(\sigma_n x_n, \varepsilon_n) \end{pmatrix} \tag{5}$$

where

$$\sigma = \begin{pmatrix} \sigma_1 & 0 & \cdots & 0 \\ 0 & \sigma_2 & \cdots & 0 \\ \vdots & \vdots & \ddots & \vdots \\ 0 & 0 & \cdots & \sigma_n \end{pmatrix} \tag{6}$$

is the gain matrix, and $\varepsilon$ is an upper bound for the controller (5) which is also to be designed:

$$\varepsilon = [\varepsilon_1, \varepsilon_2, \cdots, \varepsilon_n]^T \tag{7}$$

In this paper, a unified approach to chaotifying the continuous-time autonomous system (1)-(3) is proposed in a general form, by means of designing $B$, $G(\sigma x, \varepsilon)$, $\sigma$, and $\varepsilon$ such that the controlled system (3) becomes chaotic in the sense of being globally bounded with a required placement of Lyapunov exponents.

## 3. General Principles and Design Criteria

### 3.1 Design Principles

Consider a general $n$-dimensional nonlinear autonomous system (1)-(2), as shown in Fig. 1, and the controlled system (3) shown in Fig. 2.

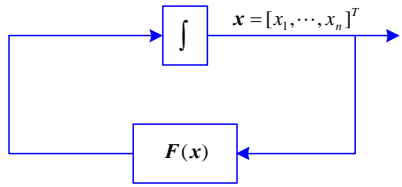 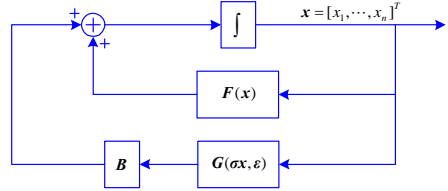

**Fig. 1** The uncontrolled system          **Fig. 2** The controlled system

Some basic principles for designing an anti-controller (5) are first discussed.

Recall that in the discrete-time setting, there are effective Chen-Lai algorithm and Wang-Chen scheme, which enforce the controlled system to have a diagonally dominant system matrix thereby yielding all positive Lyapunov exponents so that the controlled system orbits can expand in all directions, and then by designing a bounding force into the controller one can fold the outgoing orbits back to ensure the overall global boundedness. A combination of these two (expanding and folding) actions leads the controlled system to be chaotic in the sense of Li and Yorke [15-26]. For continuous-time systems, however, a chaotic system does not have all positive Lyapunov exponents. Instead, these exponents have to be placed in a certain particular pattern. For 3-dimensional autonomous chaotic systems, for example, their three Lyapunov exponents always have the three different signs of $(+,-,0)$.

It should be noted that, for nonlinear systems, even if the eiganvalues of their Jacobians at equilibria are all located on the left-half complex plane, a uniformly bounded nonlinear controller may not be able to ensure the controlled system be globally bounded.

To overcome the above-mentioned technical difficulties, for a system consisting of a linear and a nonlinear subsystem, a new chaotification method is developed here using separation of state variables to decompose the system into two open-loop subsystems which are interacted via mutual feedback, so as to result in an overall closed-loop nonlinear feedback control system. The new method has the following five aspects to consider in the controller design:

1) Assume that the given system (1) is composed of a linear and a nonlinear subsystem, or otherwise the nonlinear controller can be so designed to result in such decomposition, namely, in the following form:

$$\begin{cases} \dot{x}_{1,m} = A_{11}x_{1,m} + f(x_{m+1,n}) \\ \dot{x}_{m+1,n} = A_{21}x_{1,m} + A_{22}x_{m+1,n} \end{cases} \quad (8)$$

Here, $x_{1,m} = [x_1, x_2, \cdots, x_m]^T$, $x_{m+1,n} = [x_{m+1}, x_{m+2}, \cdots, x_n]^T$, $x = [x_{1,m}, x_{m+1,n}]^T$, and $A_{11}x_{1,m}$, $A_{21}x_{1,m}$ and $A_{22}x_{m+1,n}$ are the linear parts of $F(x)$ in the forms of

$$\begin{cases} A_{11}x_{1,m} = \begin{pmatrix} a_{11} & a_{12} & \cdots & a_{1m} \\ a_{21} & a_{22} & \cdots & a_{2m} \\ \vdots & \vdots & \ddots & \vdots \\ a_{m1} & a_{m2} & \cdots & a_{mm} \end{pmatrix} \begin{pmatrix} x_1 \\ x_2 \\ \vdots \\ x_m \end{pmatrix}, \quad A_{21}x_{1,m} = \begin{pmatrix} a_{m+1,1} & a_{m+1,2} & \cdots & a_{m+1,m} \\ a_{m+2,1} & a_{m+2,2} & \cdots & a_{m+2,m} \\ \vdots & \vdots & \ddots & \vdots \\ a_{n1} & a_{n2} & \cdots & a_{nm} \end{pmatrix} \begin{pmatrix} x_1 \\ x_2 \\ \vdots \\ x_m \end{pmatrix} \\ A_{22}x_{m+1,n} = \begin{pmatrix} a_{m+1,m+1} & a_{m+1,m+2} & \cdots & a_{m+1,n} \\ a_{m+2,m+1} & a_{m+2,m+2} & \cdots & a_{m+2,n} \\ \vdots & \vdots & \ddots & \vdots \\ a_{n,m+1} & a_{n,m+2} & \cdots & a_{nn} \end{pmatrix} \begin{pmatrix} x_{m+1} \\ x_{m+2} \\ \vdots \\ x_n \end{pmatrix} \end{cases} \quad (9)$$

Moreover, $f(x_{m+1,n})$ is the nonlinear part of $F(x)$, in the form of

$$f(x_{m+1,n}) = \begin{pmatrix} f_1(x_{m+1,n}) \\ f_2(x_{m+1,n}) \\ \vdots \\ f_m(x_{m+1,n}) \end{pmatrix} \quad (10)$$

where $f_i(x_{m+1,n})$, $i = 1, 2, \ldots, m$, are arbitrary but bounded-input/bounded-output nonlinear (e.g., polynomial, exponential, logarithmic, hyperbolic, sinusoidal, signum) functions.

2) Set matrix $A_{21}$ in (9) be zero, namely,

$$A_{21} = \begin{pmatrix} 0 & 0 & \cdots & 0 \\ 0 & 0 & \cdots & 0 \\ \vdots & \vdots & \ddots & \vdots \\ 0 & 0 & \cdots & 0 \end{pmatrix}_{(n-m) \times m} \quad (11)$$

3) By separation of variables, reformulate the uncontrolled system (8) to be in the following general form:

$$\begin{cases} \dot{x}_{1,m} = A_{11}x_{1,m} + f(x_{m+1,n}) \\ \dot{x}_{m+1,n} = A_{22}x_{m+1,n} \end{cases} \quad (12)$$

This decomposes the uncontrolled system (8) into two open-loop subsystems. Specifically, if $A_{21}$ is zero, then the original state vector $x = [x_1, x_2, \cdots, x_n]^T$ is decomposed into $x_{1,m} = [x_1, x_2, \cdots, x_m]^T$ and $x_{m+1,n} = [x_{m+1}, x_{m+2}, \cdots, x_n]^T$, so the uncontrolled system shown in Fig. 1 becomes two open-loop subsystems as shown in Fig. 3.

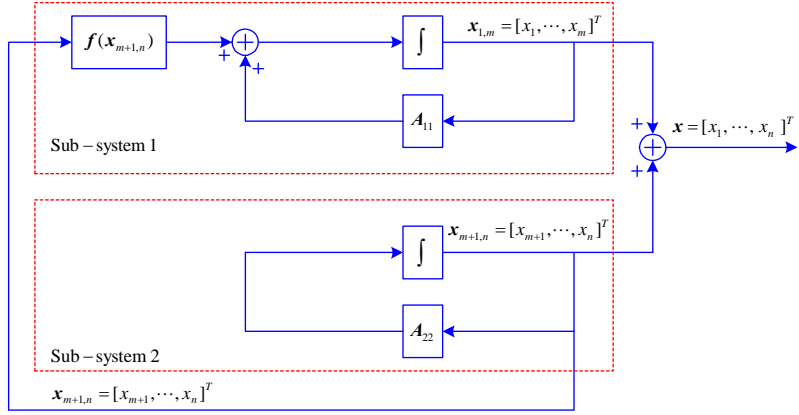

**Fig. 3** Two open-loop subsystems

4) Through the nonlinear controller $g(\sigma_{1,m}x_{1,m},\varepsilon_{1,m})$ with gain matrix $B$, both of which are to be designed, and using the output component $x_{1,m}$ from the first subsystem, couple the two subsystems together as shown in Fig. 4. This coupled system still has two subsystems but overall is a closed-loop state-feedback system.

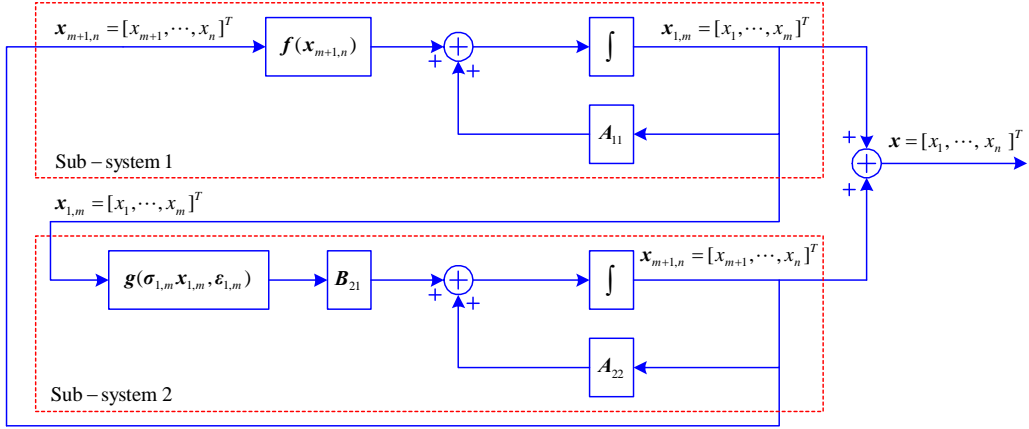

**Fig. 4** A closed-loop state-feedback controlled system

5) According to Fig. 4, one actually has obtained the following equations:

$$\begin{cases} \dot{x}_{1,m} = A_{11}x_{1,m} + f(x_{m+1,n}) \\ \dot{x}_{m+1,n} = A_{22}x_{m+1,n} + B_{21}g(\sigma_{1,m}x_{1,m},\varepsilon_{1,m}) \end{cases} \quad (13)$$

where $B_{21}$ is a sub-matrix of the control matrix $B$, in the form of

$$B = \begin{pmatrix} b_{11} & b_{12} & \cdots & b_{1n} \\ b_{21} & b_{22} & \cdots & b_{2n} \\ \vdots & \vdots & \ddots & \vdots \\ b_{n1} & b_{n2} & \cdots & b_{nn} \end{pmatrix} = \begin{pmatrix} B_{11} & B_{12} \\ B_{21} & B_{22} \end{pmatrix}_{n \times n} \quad (14)$$

in which

$$\begin{cases} \boldsymbol{B}_{11} = \begin{pmatrix} 0 & 0 & \cdots & 0 \\ 0 & 0 & \cdots & 0 \\ \vdots & \vdots & \ddots & \vdots \\ 0 & 0 & \cdots & 0 \end{pmatrix}_{m \times m} & \boldsymbol{B}_{12} = \begin{pmatrix} 0 & 0 & \cdots & 0 \\ 0 & 0 & \cdots & 0 \\ \vdots & \vdots & \ddots & \vdots \\ 0 & 0 & \cdots & 0 \end{pmatrix}_{m \times (n-m)} \\ \boldsymbol{B}_{21} = \begin{pmatrix} b_{m+1,1} & b_{m+1,2} & \cdots & b_{m+1,m} \\ b_{m+2,1} & b_{m+2,2} & \cdots & b_{m+2,m} \\ \vdots & \vdots & \ddots & \vdots \\ b_{n1} & b_{n2} & \cdots & b_{nm} \end{pmatrix}_{(n-m) \times m} & \boldsymbol{B}_{22} = \begin{pmatrix} 0 & 0 & \cdots & 0 \\ 0 & 0 & \cdots & 0 \\ \vdots & \vdots & \ddots & \vdots \\ 0 & 0 & \cdots & 0 \end{pmatrix}_{(n-m) \times (n-m)} \end{cases} \quad (15)$$

And the nonlinear controller $\boldsymbol{g}(\boldsymbol{\sigma}_{1,m}\boldsymbol{x}_{1,m},\boldsymbol{\varepsilon}_{1,m})$ has the form of

$$\boldsymbol{g}(\boldsymbol{\sigma}_{1,m}\boldsymbol{x}_{1,m},\boldsymbol{\varepsilon}_{1,m}) = \begin{pmatrix} g_1(\sigma_1 x_1, \varepsilon_1) \\ g_2(\sigma_2 x_2, \varepsilon_2) \\ \vdots \\ g_m(\sigma_m x_m, \varepsilon_m) \end{pmatrix} \quad (16)$$

with a gain matrix $\boldsymbol{\sigma}_{1,m}$ and control output upper bound $\boldsymbol{\varepsilon}_{1,m}$ given by

$$\begin{cases} \boldsymbol{\sigma}_{1,m} = \begin{pmatrix} \sigma_1 & 0 & \cdots & 0 \\ 0 & \sigma_2 & \cdots & 0 \\ \vdots & \vdots & \ddots & \vdots \\ 0 & 0 & \cdots & \sigma_m \end{pmatrix} \\ \boldsymbol{\varepsilon}_{1,m} = [\varepsilon_1, \varepsilon_2, \cdots, \varepsilon_m]^T \end{cases} \quad (17)$$

Note that in (8)-(17), $m = 1, 2, \cdots, n-1$, and all matrix entries are real constants.

### 3.2 Theoretical Analysis

The following is the main theoretical result of the paper.

**Theorem 1** Assume that all the eigenvalues of the sub-matrices $A_{11}$ and $A_{22}$ in system (13) have negative real parts, the nonlinear controller $\boldsymbol{g}(\boldsymbol{\sigma}_{1,m}\boldsymbol{x}_{1,m},\boldsymbol{\varepsilon}_{1,m})$ is uniformly bounded:

$$\sup_{0 \leq t < \infty} \| \boldsymbol{g}(\boldsymbol{\sigma}_{1,m}\boldsymbol{x}_{1,m},\boldsymbol{\varepsilon}_{1,m}) \| \leq \| \boldsymbol{\varepsilon}_{1,m} \| < \infty \quad (18)$$

and $\boldsymbol{f}(\boldsymbol{x}_{m+1,n})$ is a bounded-input/bounded-output function, namely satisfying

$$\| \boldsymbol{f}(\boldsymbol{x}_{m+1,n}) \| \leq M_1 < \infty \quad (\text{if } \| \boldsymbol{x}_{m+1,n} \| \leq M_2 < \infty) \quad (19)$$

where $\| \cdot \|$ is the Euclidean norm. Then, the controlled system (13) is globally bounded.

*Proof.* First, it is to prove that the solution orbit of the second equation in (13) is globally bounded. Indeed, it is well-known that this solution is given by

$$\boldsymbol{x}_{m+1,n}(t) = \exp(\boldsymbol{A}_{22}t) \cdot \boldsymbol{x}_{m+1,n}(0) + \int_0^t \exp[\boldsymbol{A}_{22}(t-\tau)] \cdot \boldsymbol{B}_{21}\boldsymbol{g}(\boldsymbol{\sigma}_{1,m}\boldsymbol{x}_{1,m},\boldsymbol{\varepsilon}_{1,m})d\tau \qquad (20)$$

where $\boldsymbol{x}_{m+1,n}(t) = [x_{m+1}(t), x_{m+2}(t), \cdots, x_n(t)]^T$. Since all eigenvalues of the sub-matrix $\boldsymbol{A}_{22}$ have negative real parts, there exist constants $\alpha_1, \beta_1 > 0$ such that

$$\sup_{0 \le t < \infty} \| \exp(\boldsymbol{A}_{22}t) \| \le \alpha_1 e^{-\beta_1 t}$$

Furthermore, since $\sup_{0 \le t < \infty} \| \boldsymbol{g}(\boldsymbol{\sigma}_{1,m}\boldsymbol{x}_{1,m},\boldsymbol{\varepsilon}_{1,m}) \| \le \| \boldsymbol{\varepsilon}_{1,m} \| < \infty$, it follows from (20) that, for all $0 \le t < \infty$, one has

$$\begin{aligned}
\sup_{0 \le t < \infty} \| \boldsymbol{x}_{m+1,n}(t) \| &\le \sup_{0 \le t < \infty} \alpha_1 e^{-\beta_1 t} \cdot \| \boldsymbol{x}_{m+1,n}(0) \| + \sup_{0 \le t < \infty} \alpha_1 \cdot \| \boldsymbol{B}_{21} \| \cdot \| \boldsymbol{\varepsilon}_{1,m} \| \int_0^t e^{-\beta_1(t-\tau)} d\tau \\
&= \sup_{0 \le t < \infty} \alpha_1 e^{-\beta_1 t} \cdot \| \boldsymbol{x}_{m+1,n}(0) \| + \sup_{0 \le t < \infty} \frac{\alpha_1 \cdot \| \boldsymbol{B}_{21} \| \cdot \| \boldsymbol{\varepsilon}_{1,m} \|}{\beta_1}(1-e^{-\beta_1 t}) \\
&\le \alpha_1 \cdot \| \boldsymbol{x}_{m+1,n}(0) \| + \frac{\alpha_1 \cdot \| \boldsymbol{B}_{21} \| \cdot \| \boldsymbol{\varepsilon}_{1,m} \|}{\beta_1} < \infty
\end{aligned} \qquad (21)$$

Next, it is to prove that the first equation of system (13) is also globally bounded. In fact, since all eigenvalues of the sub-matrix $\boldsymbol{A}_{11}$ have negative real parts, by the boundedness of $\boldsymbol{x}_{m+1,n}$, proved above, and due to the bounded-input/bounded-output property of the nonlinear function $\boldsymbol{f}(\boldsymbol{x}_{m+1,n})$, the outputs of $\boldsymbol{f}(\boldsymbol{x}_{m+1,n})$ are all bounded. It thus follows similarly that the first equation of system (13) is globally bounded. Combining the above two cases completes the proof of the theorem.

### 3.3 Design Criteria

Based on the global boundedness established in the previous subsection, by suitably designing the parametric sub-matrices $\boldsymbol{A}_{11}$ and $\boldsymbol{A}_{22}$, the nonlinear function $\boldsymbol{f}(\boldsymbol{x}_{m+1,n})$, the control gain sub-matrix $\boldsymbol{B}_{21}$, the nonlinear controller $\boldsymbol{g}(\boldsymbol{\sigma}_{1,m}\boldsymbol{x}_{1,m},\boldsymbol{\varepsilon}_{1,m})$ as well as its gain matrix $\boldsymbol{\sigma}_{1,m}$ and the output upper bound $\boldsymbol{\varepsilon}_{1,m}$, it is possible to place the desirable positive-zero-negative Lyapunov exponents, so as to guarantee system (13) be globally bounded while possessing the desired placement of Lyapunov exponents.

**Theorem 2** (Sufficient conditions) Assume that system (13) satisfies the following six conditions. Then, the system is chaotic in the sense of having positive-zero-negative Lyaponov exponents while being globally bounded.

1) All eiganvalues of the two sub-matrices $\boldsymbol{A}_{11}$ and $\boldsymbol{A}_{22}$ have negative real parts.

2) The nonlinear controller $g(\sigma_{1,m}x_{1,m},\varepsilon_{1,m})$ is uniformly bounded:

$$\sup_{0\leq t<\infty} \| g(\sigma_{1,m}x_{1,m},\varepsilon_{1,m}) \| \leq \| \varepsilon_{1,m} \| < \infty$$

3) The nonlinear function $f(x_{m+1,n})$ is of bounded-input/bounded-output:

$$\| f(x_{m+1,n}) \| \leq M_1 < \infty \quad (\text{if } \| x_{m+1,n} \| \leq M_2 < \infty)$$

4) All equilibria $Q$ of system (13) are saddle-foci.

5) The Jacobian of the controlled system at equilibrium $Q$ is

$$J_Q = \begin{pmatrix} J_{11} & J_{12} & \cdots & J_{1n} \\ J_{21} & J_{22} & \cdots & J_{2n} \\ \vdots & \vdots & \ddots & \vdots \\ J_{n1} & J_{n2} & \cdots & J_{nn} \end{pmatrix}_Q$$

$$= \begin{pmatrix} a_{11} & \cdots & a_{1m} & \frac{\partial f_1(x_{m+1,n})}{\partial x_{m+1}} & \cdots & \frac{\partial f_1(x_{m+1,n})}{\partial x_n} \\ \vdots & \ddots & \vdots & \vdots & \ddots & \vdots \\ a_{m1} & \cdots & a_{mm} & \frac{\partial f_m(x_{m+1,n})}{\partial x_{m+1}} & \cdots & \frac{\partial f_m(x_{m+1,n})}{\partial x_n} \\ b_{m+1,1}\frac{\partial g_1(\sigma_1 x_1,\varepsilon_1)}{\partial x_1} & \cdots & b_{m+1,m}\frac{\partial g_m(\sigma_m x_m,\varepsilon_m)}{\partial x_m} & a_{m+1,m+1} & \cdots & a_{m+1,n} \\ \vdots & \ddots & \vdots & \vdots & \ddots & \vdots \\ b_{n1}\frac{\partial g_1(\sigma_1 x_1,\varepsilon_1)}{\partial x_1} & \cdots & b_{nm}\frac{\partial g_m(\sigma_m x_m,\varepsilon_m)}{\partial x_m} & a_{n,m+1} & \cdots & a_{nn} \end{pmatrix}_Q \quad (22)$$

with diagonal elements satisfying

$$\sum_{i=1}^{n} J_{ii}(Q) < 0 \quad (23)$$

6) The controlled system (13) has the desired positive-zero-negative Lyapunov exponents.

*Proof.* First, it follows from conditions 1) – 3) and Theorem 1 that system (13) is globally bounded.

Then, condition 4) implies that the system has a potential to diverge; therefore, by suitably designing the parameters in $A_{11}$, $A_{22}$, $f(x_{m+1,n})$, $B_{21}$, $g(\sigma_{1,m}x_{1,m},\varepsilon_{1,m})$, $\sigma_{1,m}$, $\varepsilon_{1,m}$, it is possible to modify the real (and imaginary) parts of the Jacobian eigenvalues, so as to obtain positive Lyapunov exponents.

Furthermore, by condition 5), the system is dissipative therefore negative Luapunov exponents can be placed at will.

Finally, condition 6) guarantees to have the desired placement of all Lyapunov exponents. As a result, the desired chaotic system can be obtained.

## 4. Design Examples

### 4.1 3-D nonlinear systems

**Example 1.** Consider the following system:

$$\begin{cases} \dot{x}_{1,m} = A_{11}x_{1,m} + f(x_{m+1,n}) \\ \dot{x}_{m+1,n} = A_{22}x_{m+1,n} + B_{21}g(\sigma_{1,m}x_{1,m}, \varepsilon_{1,m}) \end{cases} \quad (24)$$

where $n = 3$, $m = 1$, and a saw-tooth function

$$g(\sigma_{1,m}x_{1,m}, \varepsilon_{1,m}) = (g_1(\sigma_1 x_1, \varepsilon_1)) = (\varepsilon_1 \cdot \text{sawtooth}(\pi\sigma_1(x_1 - \varepsilon_1/\sigma_1)/\varepsilon_1, p)) \quad (25)$$

in which $p = 1$, $\sigma_1 = 28$, $\varepsilon_1 = 14$, $\sup_{0 \le t < \infty} \|g(\sigma_{1,m}x_{1,m}, \varepsilon_{1,m})\| \le \|\varepsilon_{1,m}\| = 14 < \infty$, and $g_1(\sigma_1 x_1, \varepsilon_1)$ is visualized in Fig.5.

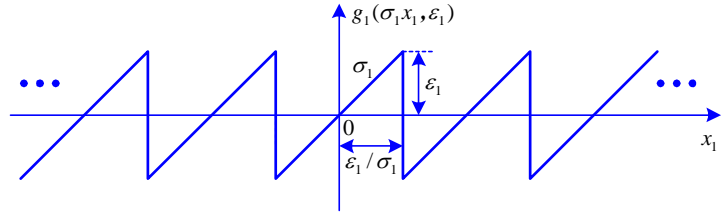

**Fig. 5** Graph of function $g_1(\sigma_1 x_1, \varepsilon_1)$

Choose the system sub-matrices $A_{11}$, $A_{22}$ and control matrix $B_{21}$ as

$$A_{11} = -1, \quad A_{22} = \begin{pmatrix} -1 & 10 \\ -5 & -1 \end{pmatrix}, \quad B_{21} = \begin{pmatrix} 1 \\ 1 \end{pmatrix}$$

It can be easily verified that the eigenvalues of $A_{11}$ and $A_{22}$ are located on the left-half complex plane. Choose the nonlinear function

$$f(x_{m+1,n}) = (-x_2 + 0.2(x_3)^2)$$

Clearly, this function is of bounded-input/bounded-output.

Now, substituting $A_{11}$, $A_{22}$, $B_{21}$ and $f(x_{m+1,n})$ into (24) yields a controlled system of the form

$$\begin{cases} \dot{x}_1 = -x_1 + f(x_{m+1,n}) \\ \dot{x}_2 = g_1(\sigma_1 x_1, \varepsilon_1) - x_2 + 10x_3 \\ \dot{x}_3 = g_1(\sigma_1 x_1, \varepsilon_1) - 5x_2 - x_3 \end{cases} \quad (26)$$

Its corresponding uncontrolled system

$$\begin{cases} \dot{x}_1 = -x_1 - x_2 + 0.2(x_3)^2 \\ \dot{x}_2 = -x_2 + 10x_3 \\ \dot{x}_3 = -5x_2 - x_3 \end{cases}$$

has a unique stable equilibrium, $Q_0(0,0,0)$, at which the Jacobian eigenvalues are

$\lambda_1 = -1.0$, and $\lambda_{2,3} = -1.0 \pm j7.0711$.

Note that the equilibria of the controlled system can be obtained by solving the following equations:

$$\begin{cases} -x_1^* - x_2^* + 0.2(x_3^*)^2 = 0 \\ -x_2^* + 10x_3^* + g_1(\sigma_1 x_1^*, \varepsilon_1) = 0 \\ -5x_2^* - x_3^* + g_1(\sigma_1 x_1^*, \varepsilon_1) = 0 \end{cases}$$

which gives

$$\begin{cases} x_1^* = -(11/51)g_1(\sigma_1 x_1^*, \varepsilon_1) + 0.2(4/51)^2[g_1(\sigma_1 x_1^*, \varepsilon_1)]^2 \\ x_2^* = (11/51)g_1(\sigma_1 x_1^*, \varepsilon_1) \\ x_3^* = -(4/51)g_1(\sigma_1 x_1^*, \varepsilon_1) \end{cases} \quad (27)$$

yielding

$$\begin{cases} Q_1(-2.6000, 2.4157, -0.8784) \\ Q_2(-1.7000, 1.8118, -0.6588) \\ Q_3(-0.8000, 1.2078, -0.4392) \\ Q_4(0,0,0) \\ Q_5(0.8500, -0.9059, 0.3294) \\ Q_6(1.7500, -1.5098, 0.5490) \\ Q_7(2.6500, -2.1137, 0.7686) \end{cases}$$

It follows from the first equation of (27) that, at the seven equilibria, the $x_1$ coordinates of the above seven equilibria have a distribution as shown in Fig. 6.

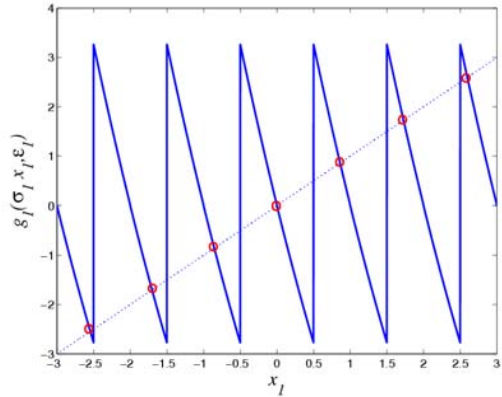

**Fig. 6** Distribution of the $x_1$ coordinates of the seven equilibria

Moreover, the Jacobian of the controlled system (26) is given by

$$J_{Q_i} = \begin{pmatrix} -1 & -1 & 0.4x_3^* \\ \sigma_1 & -1 & 10 \\ \sigma_1 & -5 & -1 \end{pmatrix}_{Q_i}$$

which has the following eigenvalues at the seven equilibria, respectively:

$$\begin{cases} Q_1: -3.4585, 0.2292 + j9.6110 \\ Q_2: -3.6335, 0.3167 + j9.5174 \\ Q_3: -3.8120, 0.4060 + j9.4260 \\ Q_4: -4.1782, 0.5891 + j9.2507 \\ Q_5: -4.4592, 0.7296 + j9.1261 \\ Q_6: -4.6488, 0.8244 + j9.0463 \\ Q_7: -4.8396, 0.9198 + j8.9693 \end{cases}$$

These are all saddle-nodes of index 2. Thus, all conditions in Theorem 2 are satisfied, so the controlled system is chaotic. In face, its attractor is obtained as shown in Fig. 7.

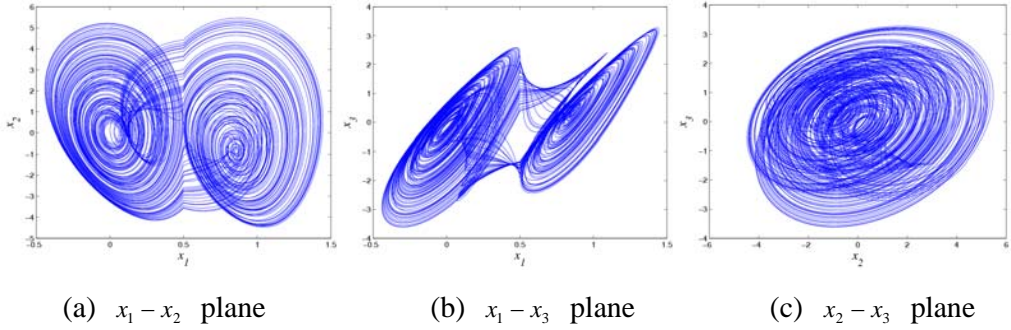

(a) $x_1 - x_2$ plane  (b) $x_1 - x_3$ plane  (c) $x_2 - x_3$ plane

**Fig. 7** Chaotic attractor of Example 1

**Example 2.** Assume that the controlled system is same as (26), in which all parameters remain unchanged, but equipped with an exponential function given by

$$f(\boldsymbol{x}_{m+1,n}) = \left( -x_2 + \frac{x_3}{1 + e^{(x_3)^2}} \right)$$

It is obvious that if $x_2, x_3$ are bounded, then the output of $f(\boldsymbol{x}_{m+1,n})$ is also bounded. Similarly, all conditions in Theorem 2 are satisfied, so that the controlled system is chaotic, with an attractor as shown in Fig. 8.

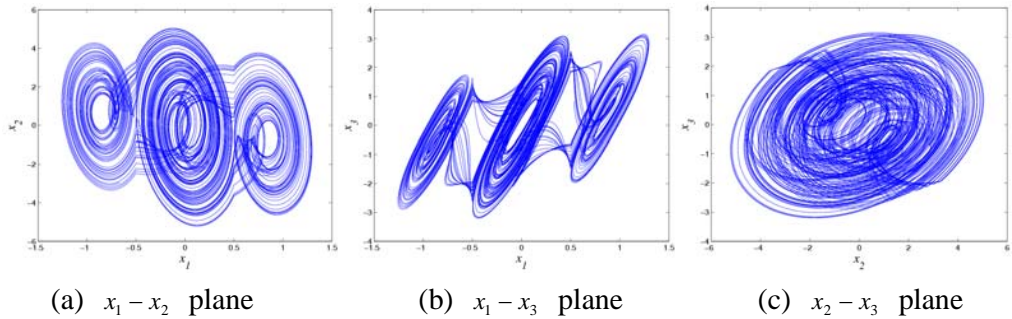

(a) $x_1 - x_2$ plane  (b) $x_1 - x_3$ plane  (c) $x_2 - x_3$ plane

**Fig. 8** Chaotic attractor of Example 2

**Example 3.** Assume that the controlled system is same as (26), where all parameters remain unchanged, but equipped with an exponential function given by

$$f(\boldsymbol{x}_{m+1,n}) = (-x_2 + \mathrm{sgn}(x_2))$$

It is obvious that if $x_2, x_3$ are bounded, then the output of $f(\boldsymbol{x}_{m+1,n})$ is also bounded. Similarly, all conditions in Theorem 2 are satisfied, therefore the controlled system is chaotic, with an attractor as shown in Fig. 9.

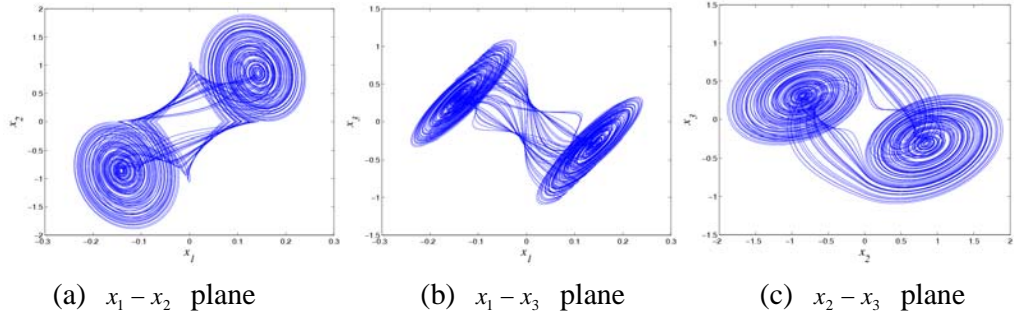

(a) $x_1 - x_2$ plane  (b) $x_1 - x_3$ plane  (c) $x_2 - x_3$ plane

**Fig. 9** Chaotic attractor of Example 3

### 4.2 A 4-D nonlinear system

**Example 4.** Consider a 4-D nonlinear system with state variables $x_1, x_2, x_3$ used for control. Its general form of state equations are

$$\begin{cases} \dot{\boldsymbol{x}}_{1,m} = \boldsymbol{A}_{11}\boldsymbol{x}_{1,m} + \boldsymbol{f}(\boldsymbol{x}_{m+1,n}) \\ \dot{\boldsymbol{x}}_{m+1,n} = \boldsymbol{A}_{22}\boldsymbol{x}_{m+1,n} + \boldsymbol{B}_{21}\boldsymbol{g}(\boldsymbol{\sigma}_{1,m}\boldsymbol{x}_{1,m}, \boldsymbol{\varepsilon}_{1,m}) \end{cases} \quad (28)$$

where $n = 4, m = 3$, along with a sine function

$$\boldsymbol{g}(\boldsymbol{\sigma}_{1,m}\boldsymbol{x}_{1,m}, \boldsymbol{\varepsilon}_{1,m}) = \begin{pmatrix} g_1(\sigma_1 x_1, \varepsilon_1) \\ g_2(\sigma_2 x_2, \varepsilon_2) \\ g_3(\sigma_3 x_3, \varepsilon_3) \end{pmatrix} = \begin{pmatrix} \varepsilon_1 \cdot \sin(\sigma_1 x_1) \\ \varepsilon_2 \cdot \sin(\sigma_2 x_2) \\ \varepsilon_3 \cdot \sin(\sigma_3 x_3) \end{pmatrix} \quad (29)$$

in which $\sigma_1 = 2$, $\sigma_2 = 4$, $\sigma_3 = 6$, $\varepsilon_1 = 0.5$, $\varepsilon_2 = 1.0$, $\varepsilon_3 = 1.5$, and $\sup_{0 \le t < \infty} \|\boldsymbol{g}(\boldsymbol{\sigma}_{1,m}\boldsymbol{x}_{1,m}, \boldsymbol{\varepsilon}_{1,m})\| \le 1.5 < \infty$.

Choose the system sub-matrices $\boldsymbol{A}_{11}$, $\boldsymbol{A}_{22}$ and control matrix $\boldsymbol{B}_{21}$ as

$$\boldsymbol{A}_{11} = \begin{pmatrix} -0.5 & -4.9 & 5.1 \\ 4.9 & -5.3 & 0.1 \\ -5.1 & 0.1 & 4.7 \end{pmatrix}, \quad \boldsymbol{A}_{22} = -1, \quad \boldsymbol{B}_{21} = (1 \ 1 \ 1) \quad (30)$$

It can be verified that the eigenvalues of $\boldsymbol{A}_{11}$ and $\boldsymbol{A}_{22}$ are located on the left-half complex plane. Choose the nonlinear function

$$\boldsymbol{f}(\boldsymbol{x}_{m+1,n}) = \begin{pmatrix} f_1(\boldsymbol{x}_{m+1,n}) \\ f_2(\boldsymbol{x}_{m+1,n}) \\ f_3(\boldsymbol{x}_{m+1,n}) \end{pmatrix} = \begin{pmatrix} x_4 + (x_4)^2 \\ x_4 \\ -x_4 + (x_4)^2 \end{pmatrix} \quad (31)$$

Clearly this function is of bounded-input/bounded-output.

Now, substituting $A_{11}$, $A_{22}$, $B_{21}$, $\boldsymbol{g}(\sigma_{1,m}\boldsymbol{x}_{1,m},\varepsilon_{1,m})$, and $\boldsymbol{f}(\boldsymbol{x}_{m+1,n})$ into (28) yields a controlled system in the form of

$$\begin{cases} \dot{x}_1 = -0.5x_1 - 4.9x_2 + 5.1x_3 + f_1(\boldsymbol{x}_{m+1,n}) \\ \dot{x}_2 = 4.9x_1 - 5.3x_2 + 0.1x_3 + f_2(\boldsymbol{x}_{m+1,n}) \\ \dot{x}_3 = -5.1x_1 + 0.1x_2 + 4.7x_3 + f_3(\boldsymbol{x}_{m+1,n}) \\ \dot{x}_4 = g_1(\sigma_1 x_1,\varepsilon_1) + g_2(\sigma_2 x_2,\varepsilon_2) + g_3(\sigma_3 x_3,\varepsilon_3) - x_4 \end{cases} \quad (32)$$

Similarly, all conditions in Theorem 2 are satisfied, so the controlled system is chaotic, with an attractor as shown in Fig. 10.

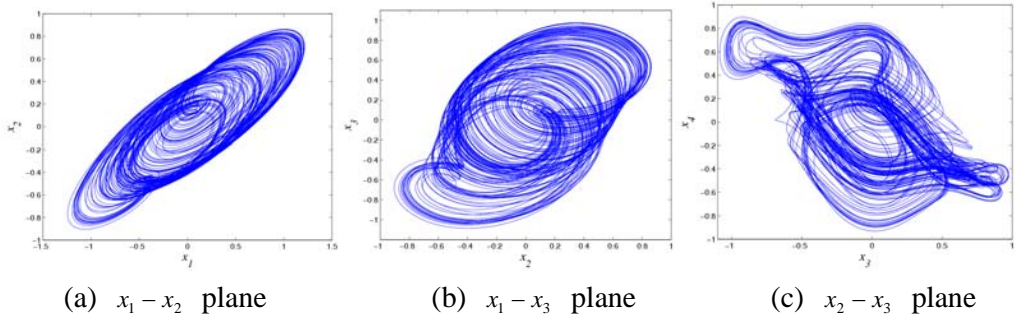

(a) $x_1 - x_2$ plane  (b) $x_1 - x_3$ plane  (c) $x_2 - x_3$ plane

**Fig. 10** Chaotic attractor of Example 4

### 4.3 A 5-D nonlinear system

**Example 5.** Consider a 5-D nonlinear system with state variables $x_1, x_2$ used for control. Its general form of state equations is

$$\begin{cases} \dot{\boldsymbol{x}}_{1,m} = A_{11}\boldsymbol{x}_{1,m} + \boldsymbol{f}(\boldsymbol{x}_{m+1,n}) \\ \dot{\boldsymbol{x}}_{m+1,n} = A_{22}\boldsymbol{x}_{m+1,n} + B_{21}\boldsymbol{g}(\sigma_{1,m}\boldsymbol{x}_{1,m},\varepsilon_{1,m}) \end{cases} \quad (33)$$

where $n = 5, m = 2$, along with a mode function

$$\boldsymbol{g}(\sigma_{1,m}\boldsymbol{x}_{1,m},\varepsilon_{1,m}) = \begin{pmatrix} g_1(\sigma_1 x_1,\varepsilon_1) \\ g_2(\sigma_2 x_2,\varepsilon_2) \end{pmatrix} = \begin{pmatrix} \mathrm{mod}(\sigma_1 x_1,\varepsilon_1) \\ \mathrm{mod}(\sigma_2 x_2,\varepsilon_2) \end{pmatrix} \quad (34)$$

in which $\sigma_1 = 10$, $\sigma_2 = 15$, $\varepsilon_1 = 5$, $\varepsilon_2 = 7.5$, and $\sup_{0 \leq t < \infty}\|\boldsymbol{g}(\sigma_{1,m}\boldsymbol{x}_{1,m},\varepsilon_{1,m})\| \leq 7.5 < \infty$.

Choose the system sub-matrices $A_{11}$, $A_{22}$ and control matrix $B_{21}$ as

$$A_{11} = \begin{pmatrix} -1 & 0 \\ 0 & -1 \end{pmatrix}, \quad A_{22} = \begin{pmatrix} -0.5 & -4.9 & 5.1 \\ 4.9 & -5.3 & 0.1 \\ -5.1 & 0.1 & 4.7 \end{pmatrix}, \quad B_{21} = \begin{pmatrix} 1 & -1 \\ 1 & 1 \\ 1 & -1 \end{pmatrix} \qquad (35)$$

It can be verified that the eigenvalues of $A_{11}$ and $A_{22}$ are located on the left-half complex plane. Choose the nonlinear function to be a polynomial:

$$f(x_{m+1,n}) = \begin{pmatrix} f_1(x_{m+1,n}) \\ f_2(x_{m+1,n}) \end{pmatrix} = \begin{pmatrix} x_3 + x_4 + x_5 - x_4 x_5 \\ x_3 + x_4 - x_5 \end{pmatrix} \qquad (36)$$

Clearly, this function is of bounded-input/bounded-output.

Now, substituting $A_{11}$, $A_{22}$, $B_{21}$, $g(\sigma_{1,m}x_{1,m}, \varepsilon_{1,m})$, and $f(x_{m+1,n})$ into (33) yields a controlled system in the form of

$$\begin{cases} \dot{x}_1 = -x_1 + f_1(x_{m+1,n}) \\ \dot{x}_2 = -x_2 + f_2(x_{m+1,n}) \\ \dot{x}_3 = g_1(\sigma_1 x_1, \varepsilon_1) - g_2(\sigma_2 x_2, \varepsilon_2) - 0.5 x_3 - 4.9 x_4 + 5.1 x_5 \\ \dot{x}_4 = g_1(\sigma_1 x_1, \varepsilon_1) + g_2(\sigma_2 x_2, \varepsilon_2) + 4.9 x_3 - 5.3 x_4 + 0.1 x_5 \\ \dot{x}_5 = g_1(\sigma_1 x_1, \varepsilon_1) - g_2(\sigma_2 x_2, \varepsilon_2) - 5.1 x_3 + 0.1 x_4 + 4.7 x_5 \end{cases} \qquad (37)$$

Similarly, all conditions in Theorem 2 are satisfied, thus the controlled system is chaotic, with an attractor as shown in Fig. 11.

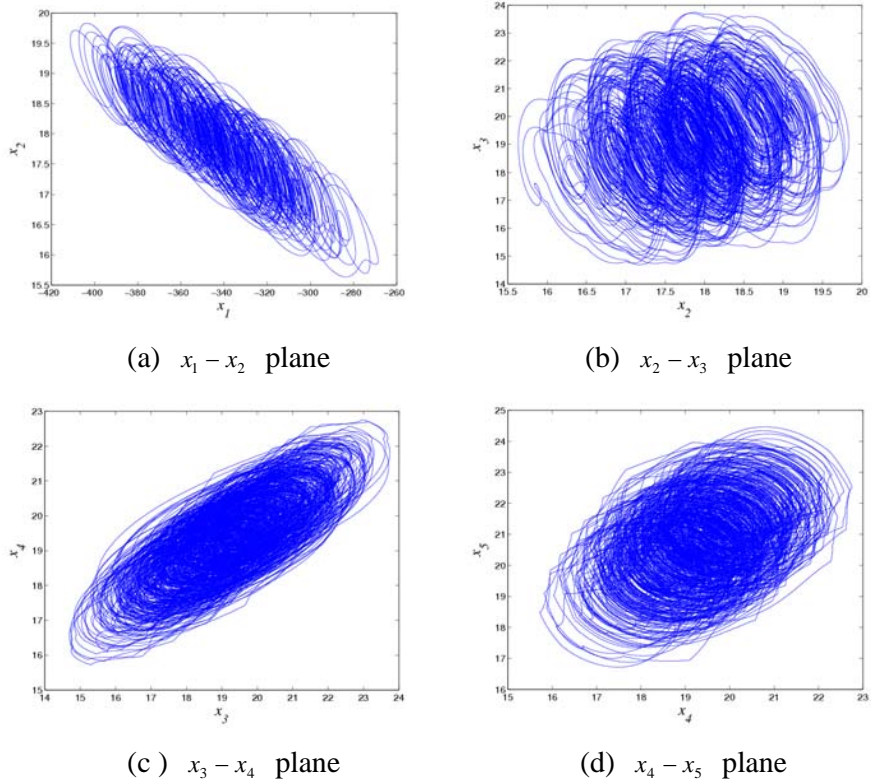

(a) $x_1 - x_2$ plane  (b) $x_2 - x_3$ plane

(c) $x_3 - x_4$ plane  (d) $x_4 - x_5$ plane

**Fig. 11** Chaotic attractor of Example 5

## 5. Conclusions

This paper has developed an effective and unified chaotification approach for designing an anti-controller which can make a general continuous-time nonlinear autonomous system chaotic in the sense of having desired positive-zero-negative Lyapunov exponents while being globally bounded. For a system consisting of a linear and a nonlinear subsystem, or otherwise by designing a suitable combined linear-nonlinear controller to obtain this structure, chaotification can be achieved. By using separation of state variables, which decomposes the system into two open-loop subsystems which are interacted through mutual feedback, an overall closed-loop nonlinear feedback control system is obtained. To that end, under the condition that the nonlinear feedback control output is uniformly bounded where the nonlinear function is of bounded-input/bounded-output, chaotification can be accomplished following a few guidelines. Furthermore, it has been theoretically proved that the resulting system is chaotic in the sense described above. Finally, several numerical examples have been given to verify the effectiveness of the theoretical design.

It should be remarked that since linear systems are special cases of nonlinear systems, the new method developed in this paper is also applicable to linear systems in general.

## Acknowledgments

This work was supported by the National Natural Science Foundation of China under Grants 61172023 and 60871025, the Natural Science Foundation of Guangdong Province under Grants 8151009001000060 and S2011010001018, the Specialized Research Foundation of Doctoral Subject Point of Education Ministry under Grant 201144420110003, and by the Hong Kong Research Grants Council under the RGF Grant CityU1114/11E.

## References

[1] Wang X F. Generating chaos in continuous-time systems via feedback control. In Chaos Control: Theory and Applications (Chen G, Yu X H, Eds), Springer-Verlag, Berlin, pp. 179-204, 2003

[2] Wang X F, Chen G R. Chaotifying a stable LTI system by tiny feedback control. IEEE Trans. Circuits Syst. I, 2000, 47: 410-415

[3] Wang X F, Chen G R, Yu X H. Anticontrol of chaos in continuous-time systems via time-delay feedback. Chaos, 2000, 10(4): 771-779

[4] Zhou T, Chen G, Yang Q. A simple time-delay feedback anti-control method made rigorous. Chaos, 2004, 14(3): 662-668

[5] Wang X F, Chen G R. Generating topologically conjugate chaotic systems via feedback control. IEEE Trans. Circuits Syst. I, 2003, 50(6): 812-817

[6] Yang L, Liu Z R, Chen G R. Chaotifying a continuous-time system via impulsive input. Int. J. Bifur. Chaos, 2002, 12: 1121-1128

[7] Yang X S, Li Q. Chaotic attractor in a simple switching control system. Int. J. Bifur. Chaos, 2002, 12: 2255-2256

[8] Lü J, Zhou T S, Chen G R, Yang X S. Generating chaos with a switching piecewise-linear controller. Chaos, 2002, 12: 344-349

[9] Li Y X, Chen G R, Tang W K S. Controlling a unified chaotic system to hyperchaotic. IEEE Trans. Circuits Syst. II, 2005, 52: 204-207

[10] Li Y X, Tang W K S, Chen G R. Hyperchaotic evolved from the generalized Lorenz system. Int. J. Circ. Theor. Appl., 2005, 33: 235-251.

[11] Yu S M, Chen G R. Anti-control of continuous-time dynamical systems. Commun. Nonlinear Sci. Numer. Simul., 2012, 17: 2617-2627.

[12] Li Y X, Tang W K S, Chen G R. Generating hyperchaos via state feedback control. Int. J. Bifur. Chaos, 2005, 15: 3367-3375.

[13] Chen A M, Lu J A, Lü J H, Yu S M. Generating hyperchaotic Lü attractor via state feedback control. Physica A, 2006, 364: 103-110

[14] Hu G. Generating hyperchaotic attractors with three positive Lyapunov exponents via state feedback control, Int. J. Bifur. Chaos, 2009, 19(2): 651-660

[15] Chen G R. Chaotification via feedback: The discrete case. In Chaos Control: Theory and Applications, Springer-Verlag, Berlin, pp. 159-178, 2003

[16] Chen G R, Lai D J. Feedback control of Lyapunov exponents for discrete-time dynamical systems. Int. J. Bifur. Chaos, 1996, 6: 1341-1349

[17] Chen G R, Lai D J. Feedback anticontrol of chaos. Int. J. Bifur. Chaos, 1998, 8: 1585-1590

[18] Lai D J, Chen G R. Making a discrete dynamical system chaotic: Theoretical results and numerical simulations. Int. J. Bifur. Chaos, 2003, 13(11): 3437-3442

[19] Lai D J, Chen G R. Chaotification of discrete-time dynamical systems: An extension of the Chen-Lai algorithm. Int. J. Bifur. Chaos, 2004, 15(1): 109-117

[20] Wang X F, Chen G R. On feedback anticontrol of chaos. Int. J. Bifur. Chaos, 1999, 9: 1435-1441

[21] Wang X F, Chen G R. Chaotification via arbitrarily small feedback controls. Int. J. Bifur. Chaos, 2000, 10: 549-570

[22] Wang X F, Chen G R. Chaoitfying a stable map via smooth small-amplitude high-frequency feedback control. Int. J. Circ. Theor. Appl., 2000, 28: 305-312

[23] Chen G R, Shi Y M. Introduction to anti-control of discrete chaos: theory and applications. Phil. Trans. R. Soc. A, 2006, 364: 2433-2447

[24] Shi Y M, Chen G R. Chaos of discrete dynamical systems in complete metric spaces. Chaos, Solitons & Fractals, 2004, 22: 555-571

[25] Shi Y M, Chen G R. Discrete chaos in Banach spaces. Science in China, Series A Math., 2005, 48(2): 222-238

[26] Shi Y M, Yu P, Chen G R. Chaoticfication of discrete dynamical systems in Banach spaces. Int. J. Bifur. Chaos, 2006, 16(9): 2615-2636